\documentstyle[preprint,aps]{revtex}
\begin{document}
\draft
\title{The onset of exciton absorption in modulation doped GaAs quantum wells}
\author{G. Yusa, H. Shtrikman, and I. Bar-Joseph}
\address{Department of Condensed Matter Physics, The Weizmann Institute of Science,\\
Rehovot 76100, Israel}
\maketitle

\begin{abstract}
We study the evolution of the absorption spectrum of a modulation doped
GaAs/AlGaAs semiconductor quantum well with decreasing the carrier density.
We find that there is a critical density which marks the transition from a
Fermi edge singularity to a hydrogen-like behavior. At this density both the
lineshape and the transitions energies of the excitons change. We study the
density dependence of the singularity exponent $\alpha $ and show that
disorder plays an important role in determining the energy scale over which
it grows.
\end{abstract}


\narrowtext

The absorption spectrum in the presence of a Fermi sea of electrons has been
a subject of theoretical and experimental research for more than three
decades. The interest in this problem was triggered by the pioneering work
of Mahan\cite{Mahan}, who showed that in metals and bulk semiconductors this
spectrum should exhibit a power law singularity at the Fermi energy. This
singularity, which became known as the Fermi edge singularity (FES),
reflects the response of the Fermi sea electrons to the attractive potential
of the valence band hole. Mahan's work was followed by a bulk of theoretical
works, which established the many body nature of the FES and provided the
tools to treat it \cite{Mahan book}.

The FES was indeed observed in the X-ray absorption of metals but was never
observed in bulk semiconductors. Its first observation in semiconductors was
in modulation doped quantum wells. It is manifested in these systems as a
pronounced enhancement of the absorption at the Fermi edge, with an
asymmetrical lineshape: a fast rise at the low energy side and a slow fall
at high energies. A signature of the FES is its strong dependence on
temperature and electron density. The high energy slope becomes more and
more gradual when these parameters are increased \cite
{Skolnick,Livescu,Brown}.

An intriguing aspect of the behavior of a system of many electrons and a
hole is the existence of a bound state. Initially, the singularity was
associated with a bound state (the so-called Mahan exciton), which is built
out of empty conduction band states below the Fermi level, but soon after,
it was realized that this exciton is unstable \cite{Mahan book}.
Nevertheless, a bound state of electrons and a hole should exist in a
two-dimensional (2D) system. It is well known that in 2D bound states occur
for arbitrarily small attractive potentials. Thus, the attractive potential
of a valence hole should have a bound state even after the many-body
interactions have been accounted for.

The recent discovery of the charged exciton, X$^{-}$, in semiconductor
quantum wells, at low-electron densities \cite{Charged excitons,GlebPRL}
created a renewed interest in the problem. The X$^{-}$, which consists of
two electrons with opposite spins bound to a valence band hole, appears at
the absorption and emission spectra as a spectral line at some energy below
the neutral exciton line, X. A natural question which then arises is how
does the low density spectrum, of the X$^{-}$ and X bound states, transform
into the FES. This issue has been addressed in a number of theoretical
publications \cite{HawrylakPRB,HawrylakComment}. It was shown that an X$^{-}$
like bound state should persist to a high electron density, and be
manifested in the absorption spectrum as a spectral line with an asymmetric
singular shape of the form $(\omega -\omega _{0})^{-\alpha }$. Another
singular peak, which is exciton like, was predicted to appear at some energy
above that line. Indeed, this prediction was recently confirmed in CdTe
modulation doped quantum wells \cite{HuardPRL}, where the X-X$^{-}$ doublet
was observed up to densities of a few$\times $10$^{11}$ cm$^{-2}$. Attempts
to observe the X-X$^{-}$ doublet in the absorption spectrum of GaAs quantum
wells in the presence of a high density two-dimensional electron gas (2DEG)
have yielded inconclusive results \cite{Brown}.

In this paper we investigate the evolution of the absorption spectrum of a
single GaAs/AlGaAs quantum well with decreasing electron density, from $%
1.5\times 10^{11}$ to $1.5\times 10^{10}$ cm$^{-2}$. We tune the electron
density continuously by applying a voltage to a semi-transparent gate
relative to an ohmic contact, which is made into the 2DEG. We find that
there is a critical density which marks the transition from a FES to a
hydrogen-like behavior. At this critical density both the lineshape and the
transitions energies of the excitons exhibit a clear change: the singular
asymmetric line becomes a symmetric resonance and the energy difference
between X and X$^{-}$ becomes constant. We study the density dependence of
the singularity exponent $\alpha $ and show that the energy scale over which
it grows is determined by disorder.

The absorption spectrum is obtained using photocurrent spectroscopy.
Electrons are excited from surface states by the incident photons to
energies above the Schottkey barrier, and then drift into the doped region.
From there they tunnel into the well and are collected by the ohmic contacts
of the 2DEG, giving rise to a photocurrent in the gate circuit. This
mechanism gives rise to a photocurrent at photon energies which extend well
below the GaAs gap. At photon energies which can be absorbed in the quantum
well there is an additional contribution to the photocurrent, resulting from
electron-hole pairs created in the well. The characteristics of this
photocurrent will be discussed in details elsewhere. There are, however, two
important experimental points which should be mentioned:

- In this gated structures light illumination gives rise to some depletion
of the carrier density. However, since the capacitance of the structure is
constant, this only implies that we need to add some positive gate voltage
in order to recover the density in the dark. We have verified this behavior
over a large light intensity range.

- The photocurrent from the quantum well rides on a background signal, due
to surface state absorption. The value of this background photocurrent
depends on the sample structure and surface treatment. The spectra are
displayed after this background was subtracted.

The sample structure that we investigated consists of the following sequence
of layers: a $300$ nm-thick GaAs buffer, a superlattice with $50$ periods of 
$100$ nm Al$_{0.37}$Ga$_{0.67}$As and $30$ nm GaAs , a $20$ nm-thick GaAs
quantum well, a $50$ nm-thick Al$_{0.37}$Ga$_{0.67}$As spacer layer, Si
delta-doping of $1\times 10^{12}$ cm$^{-2}$, a $100$ nm Al$_{0.37}$Ga$%
_{0.67} $As, a $20$ nm thick n-type Al$_{0.37}$Ga$_{0.67}$As, and a GaAs cap
layer. The wafer is processed into a field-effect transistor structure, with
a $5$ nm thick semi-transparent PaAu gate. Determination of the electron
density $N $ under illumination is done by measuring the photoluminescence
(PL) spectrum as a function of magnetic field, and finding the magnetic
field values at which an integer number of Landau levels are filled. At
these magnetic fields there is an abrupt change in the PL spectrum \cite
{Nicholas}. This procedure works well up to $N\sim 8\times 10^{10}$ cm$^{-2}$%
. To obtain the lower electron densities we use the capacitance of this
structure to extrapolate the curve of $N(V_{\text{g}})$, where $V_{\text{g}}$
is the gate voltage.

The measurements are conducted in a liquid helium storage dewar at a
temperature of $4.2$ K, using an optical fiber based system. We use a
single-mode fiber at close proximity,\ $\sim 100$ $\mu $m, to illuminate the
sample. A second multi-mode fiber is set at a distance of a few mm from the
sample and collects the emitted PL. The sample is excited by a tunable
Ti-sapphire laser, which is coupled into the single-mode fiber. The
photocurrent is amplified by a sensitive current amplifier and measured
using a lock-in amplifier. We have conducted the measurements using power
levels from $10$ nW to $1$ mW. Except for a shift in the $N(V_{\text{g}})$
curve and some broadening at high power levels the results are power
independent. After an absorption measurement at a given gate voltage $V_{%
\text{g}}$, the laser is set to $1.541$ eV and the PL is measured.

Figure 1 shows a set of photocurrent and emission spectra taken for various
gate voltages. We label each spectrum with the corresponding Fermi energy E$%
_{\text{F}}=\pi $%
%TCIMACRO{\UNICODE{0x127}}%
%BeginExpansion
h\hskip-.2em\llap{\protect\rule[1.1ex]{.325em}{.1ex}}\hskip.2em%
%EndExpansion
$^{2}N/m^{\ast }$, where $N$ is the electron density and $m^{\ast }$\ is the
electron effective mass ($m^{\ast }=0.067m_{0}$). At large electron
densities (Figs. 1a) the absorption edge is a broad step which is shifted to
high energies relative to the emission. In Fig. 1b, which is taken at E$_{%
\text{F}}=2.5$ meV, we observe the formation of two broad peaks: one at the
absorption edge and the other a few meV higher. As the density is further
reduced (E$_{\text{F}}$\ $<1.1$ meV) the two broad peaks acquire an
asymmetric singular lineshape, characterized by a steep rise at low energies
and a slow fall at high energies (Fig. 1c-d). Following the notation of Ref. 
\cite{HawrylakPRB,HawrylakComment,HuardPRL} we label the low energy peak as $%
\omega _{1}$ and the high energy one as $\omega _{2}$. The singularity in $%
\omega _{2}$ increases very fast, and at E$_{\text{F}}=0.75$ meV it becomes
a symmetric resonance, the heavy-hole exciton (Fig. 1e). The $\omega _{1}$
peak exhibits less pronounced changes. As the density is reduced it becomes
weaker and evolves into the well known charged exciton, X$^{-}$ (Figs.
1e-f). A replica of that scenario appears at the light-hole exciton energy,
4 meV higher.

The evolution of the singularity is surprisingly fast. Figure 2 describes a
fit of\ a power law singularity $A(\omega )=(\omega -\omega _{0})^{-\alpha }$
to the high side energy of $\omega _{2}$ at E$_{\text{F}}=0.9$ meV. Such a
fitting procedure allows us to accurately extract the exponent $\alpha $ at
each density. The inset shows the dependence of $\alpha $ on E$_{\text{F}}$.
It is seen that $\alpha $ increases by nearly an order of magnitude (from $%
0.05$ to $0.4)$ in a very narrow range, $\Delta $E$_{\text{F}}=0.25$ meV,
which corresponds to reducing the electron density by less than $1\times
10^{10}$ cm$^{-2}$.

The exponent $\alpha $ is related to the phase shift of the electrons at the
Fermi surface, when scattering off the valence hole potential \cite
{HawrylakComment}. In that sense it measures the efficiency of the Fermi sea
electrons in screening that potential: the smaller $\alpha $\ is, the better
is the screening.\ Previous studies have reported a significantly broader
density range over which the singularity was observed \cite
{Skolnick,Livescu,Brown,HuardPRL}. For example, in Ref. \cite{Brown} a
similar change of $\alpha $ was observed on an energy scale of $\Delta $E$_{%
\text{F}}\sim 10$ meV, more than an order of magnitude larger than in our
measurements. It was argued that the energy scale over which $\alpha $
changes is related to the many body nature of the problem \cite
{Brown,HawrylakComment}. As we shall show in the following, disorder plays a
critical role and determines the energy scale at which the singularity is
observed.

Examining the PL spectra, we notice that at the density, at which $\alpha $
starts to grow the PL lineshape transforms from a broad single peak to two
narrow resonances, associated with the X and X$^{-}$. This behavior of the
PL was extensively studied by some of us \cite{GlebPRL,EytanPRL}. We have
shown that the appearance of excitons marks the onset of strong localization
of the electrons in the potential fluctuations of the ionized donors. These
donors, which are at a distance of 50 nm (the spacer width), are randomly
distributed in the plane and induce a spatially fluctuating electrostatic
potential at the 2DEG. At high electron density the 2DEG efficiently screens
the fluctuations, but as the density is reduced the screening becomes less
efficient and they grow considerably \cite{Efros}. Thus, the growth of $%
\alpha $ is related to the onset of strong disorder in the sample.

It should be noticed that this sample is of high quality. This is evidenced
by the high mobility ( $\sim 10^{6}$ cm$^{2}$V$^{-1}$s$^{-1}$), the narrow
exciton linewidth ($0.3$ meV), and the absence of a Stokes shift between the
PL and absorption ($<0.1$ meV) . Thus, at E$_{\text{F}}>1.1$ meV, where the
potential fluctuations are suppressed, our sample could be considered as
close to ideal. Nevertheless, the singularity is significantly suppressed at
that range. Only when disorder sets in, and electrons can not efficiently
screen the valence-hole potential, $\alpha $ becomes large enough to make $%
\omega _{2}$ and $\omega _{1}$ observable as singular peaks. Indeed, it is
indicated in Ref. \cite{HuardPRL} that there is strong disorder present in
the samples and the mobility is rather low. Large disorder is also present
in the narrow quantum wells studied in Ref. \cite{Brown}, as is evident by
the very broad exciton line $\sim 8$ meV. The wide range in which singular
lineshapes are observed in these works is, therefore, related to the large
disorder in the samples which were used and not to a fundamental energy
scale. The use of a gated sample in this study allows us to control the
onset and amount of disorder.

Let us now turn to examining the density dependence of the energies of the $%
\omega _{1}$ and $\omega _{2}$ peaks (Fig. 3). The data is presented for a
limited density range, where the singularity edge is clear and the peak
energy can be unambiguously determined. In Fig. 3a we show the $\omega _{1}$
and $\omega _{2}$ transition energies. It is seen that the $\omega _{2}$
peak shifts to higher energies with increasing density while the energy of $%
\omega _{1}$ remains nearly constant. In Fig. 3b we present the energy
difference 
%TCIMACRO{\UNICODE{0x127}}%
%BeginExpansion
h\hskip-.2em\llap{\protect\rule[1.1ex]{.325em}{.1ex}}\hskip.2em%
%EndExpansion
$(\omega _{1}-$ $\omega _{2})$ as a function of E$_{\text{F}}$. It is
evident that there exists a threshold density value, $N_{\text{c}}\approx
2.5\times 10^{10}$ cm$^{-2}$ (which corresponds to E$_{\text{F}}$\ $=0.85$
meV), below which 
%TCIMACRO{\UNICODE{0x127}}%
%BeginExpansion
h\hskip-.2em\llap{\protect\rule[1.1ex]{.325em}{.1ex}}\hskip.2em%
%EndExpansion
$(\omega _{2}-\omega _{1})$ is constant and equals $1.2$ meV. Only above
this threshold density the energy difference 
%TCIMACRO{\UNICODE{0x127}}%
%BeginExpansion
h\hskip-.2em\llap{\protect\rule[1.1ex]{.325em}{.1ex}}\hskip.2em%
%EndExpansion
$(\omega _{2}-\omega _{1})$ grows. This behavior is in clear contrast with
that reported in Ref. \cite{HuardPRL}, where it was argued that 
%TCIMACRO{\UNICODE{0x127}}%
%BeginExpansion
h\hskip-.2em\llap{\protect\rule[1.1ex]{.325em}{.1ex}}\hskip.2em%
%EndExpansion
$(\omega _{2}-\omega _{1})$ changes linearly with E$_{\text{F}}$, all the
way to zero density. The value of $1.2$ meV is well documented
experimentally as the X$^{-}$ binding energy in a $20$ nm GaAs quantum well 
\cite{GlebPRL,Shields}. In fact, it is the energy difference between the two
peaks in the PL spectra (Fig. 1). It should be emphasized that the optical
spectra (Figs. 1e-f) clearly show that the density is indeed changing in
this range, as evidenced by the exchange of oscillator strength between the
X and the X$^{-}$. This conclusion is supported by transport measurements
through the sample at this range, which show that the conductivity decreases
as the gate voltage becomes more negative.

To explain this behavior let us consider the energy spectrum of the system 
\cite{HawrylakPRB}, which is schematically shown at the inset of Fig. 3b ($%
\mu $ is the chemical potential and E$_{\text{c}}$ is the single particle
band gap). At the limit of infinitesimal density there are two bound states,
X and X$^{-}$, at a relatively large energy distance (the exciton binding
energy, E$_{\text{X}}$ ) below E$_{\text{c}}$. The energy separation between
the two bound states is E$_{\text{B}}$, the X$^{-}$ binding energy. On the
other hand, at the limit of high electron density there is only one, X$^{-}$%
-like, bound state. The absorption spectrum of the system was predicted to
consist of two peaks, with an energy difference given by \cite
{HawrylakPRB,HawrylakComment}

\begin{equation}
\text{%
%TCIMACRO{\UNICODE{0x127}}%
%BeginExpansion
h\hskip-.2em\llap{\protect\rule[1.1ex]{.325em}{.1ex}}\hskip.2em%
%EndExpansion
}(\omega _{2}-\omega _{1})=\mu -\varepsilon _{{\rm b}}
\end{equation}
where $\varepsilon _{{\rm b}}$ is the binding energy of the X$^{-}$ like
bound state. It is important to note that both $\mu \ $and $\varepsilon _{%
{\rm b}}$ in Eq. 1 are measured with respect to the same level, which we
take as the bottom of the conduction band at zero electron density, E$_{%
\text{c}}^{0}$. Equation 1 has a simple meaning: it describes the energy
cost for ionizing the X$^{-}$-like bound state by exciting one of the two
electrons to the chemical potential level $\mu $. However, it is clear that
this relation is valid only at the high density limit. At low densities $\mu
\sim 0$ and $\mid \varepsilon _{{\rm b}}\mid >$ E$_{\text{X}}$. Hence, Eq. 1
would imply that 
%TCIMACRO{\UNICODE{0x127}}%
%BeginExpansion
h\hskip-.2em\llap{\protect\rule[1.1ex]{.325em}{.1ex}}\hskip.2em%
%EndExpansion
$(\omega _{2}-\omega _{1})>$ E$_{\text{X}}$, and that we should observe a
huge increase in 
%TCIMACRO{\UNICODE{0x127}}%
%BeginExpansion
h\hskip-.2em\llap{\protect\rule[1.1ex]{.325em}{.1ex}}\hskip.2em%
%EndExpansion
$(\omega _{2}-\omega _{1})$ below a certain density. Thus, the use of Eq. 1
down to zero density is incorrect \cite{HuardPRL}, and consequently the
determination of the X$^{-}$ binding energy in CdTe has to be re-examined.
Our experimental findings show that below $N_{\text{c}}$ the correct
relation is 
%TCIMACRO{\UNICODE{0x127}}%
%BeginExpansion
h\hskip-.2em\llap{\protect\rule[1.1ex]{.325em}{.1ex}}\hskip.2em%
%EndExpansion
$(\omega _{2}-\omega _{1})=$ E$_{\text{B}}$. Only at large densities, where
there is only one bound state, we are at the limit covered by Eq. 1.

It is remarkable that at the threshold density $N_{\text{c}}$ the exciton
lineshape undergoes a drastic change. This change is clearly seen in Fig. 1:
the exciton lineshape is symmetric below $N_{\text{c}}$ (Fig. 1e-f) and
becomes a singular asymmetric line above it (Figs. 1b-d). The comparison
between the exciton lineshape in Figs. 1e and 1d is particularly
interesting. The difference is manifested not only in the high energy side
but also at the low energy side, which becomes steeper above $N_{\text{c}}$.
The fact that both the energy dispersion and lineshape change at $N_{\text{c}%
}$ implies that this density marks the transition from a hydrogen-like
behavior to a FES. It is interesting to note that the value of $N_{\text{c}}$%
, which is found here, is very close to that reported in electroreflectance
measurements \cite{Shields}, where quenching of the excitonic absorption was
studied.

Finally, we wish to comment on the behavior at high densities. It can be
seen at Fig. 3b that at densities above $N_{\text{c}}$ the energy separation
between the X and X$^{-}$ grows monotonically. The dashed line describes the
relation 
%TCIMACRO{\UNICODE{0x127}}%
%BeginExpansion
h\hskip-.2em\llap{\protect\rule[1.1ex]{.325em}{.1ex}}\hskip.2em%
%EndExpansion
$(\omega _{2}-\omega _{1})=$ E$_{\text{B}}$ $+$ E$_{\text{F}}$, (where E$_{%
\text{B}}=1.2$ meV). This relation results from Eq. 1 and was used to fit
the data in Ref. \cite{HuardPRL}. It is evident that all the measured points
in Fig. 3b are below this curve, but as the density gets higher the data
points are approaching the curve. Unfortunately, we can not determine the
behavior accurately at the high density limit. The X and X$^{-}$ lines
become too broad, and one can not reliably obtain their energy separation.
Nevertheless, it is obvious from our data at higher densities that the
energy separation indeed increases monotonically with density.

This research was supported by the Minerva foundation.

\figure Fig. 1: A series of the absorption (solid line) and PL (dotted)
spectra at several gate voltages. Each spectrum is labeled by the
corresponding Fermi level, E$_{\text{F}}=\pi $%
%TCIMACRO{\UNICODE{0x127}}%
%BeginExpansion
h\hskip-.2em\llap{\protect\rule[1.1ex]{.325em}{.1ex}}\hskip.2em%
%EndExpansion
$^{2}N_{{\rm s}}/m^{\ast }$. Each spectrum is displayed after subtraction of
the background signal (see text) and is normalized by the intensity of the X
peak.

\figure Fig. 2: (a) A fit of the function $A(\omega )=(\omega -\omega
_{0})^{-\alpha }$ to the exciton singularity at E$_{\text{F}}=0.9$ meV. (b)
The dependence of the exponent $\alpha $ on E$_{\text{F}}$.

\figure Fig. 3: (a) The energies of the X and X$^{-}$ transitions as a
function of E$_{\text{F}}$. (b) The solid circles describe the measured
energy difference 
%TCIMACRO{\UNICODE{0x127}}%
%BeginExpansion
h\hskip-.2em\llap{\protect\rule[1.1ex]{.325em}{.1ex}}\hskip.2em%
%EndExpansion
$(\omega _{2}-\omega _{1})$ as a function of E$_{\text{F}}$. The dashed line
describes the relation 
%TCIMACRO{\UNICODE{0x127}}%
%BeginExpansion
h\hskip-.2em\llap{\protect\rule[1.1ex]{.325em}{.1ex}}\hskip.2em%
%EndExpansion
$(\omega _{1}-$ $\omega _{2})=$ $1.2$ $+$ E$_{\text{F}}$. The inset shows
the schematically the energy spectrum of the system as a function of density.


\begin{references}
\bibitem{Comment muEB}  The large energy difference between $\mu $ and $%
\varepsilon _{{\rm b}}$ is explicitly shown in Fig. 4 of Ref. \cite
{HawrylakPRB}.
\end{references}
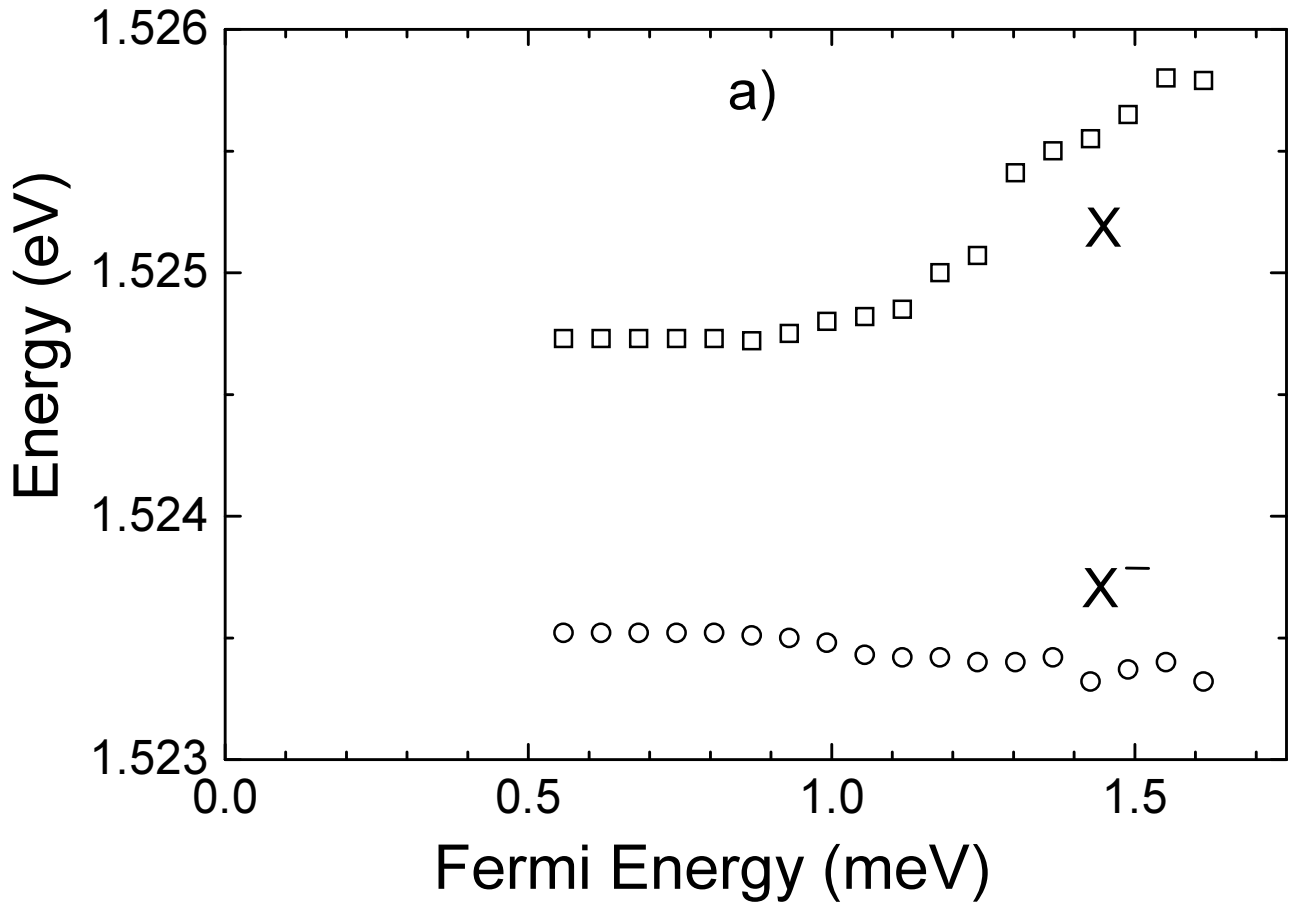
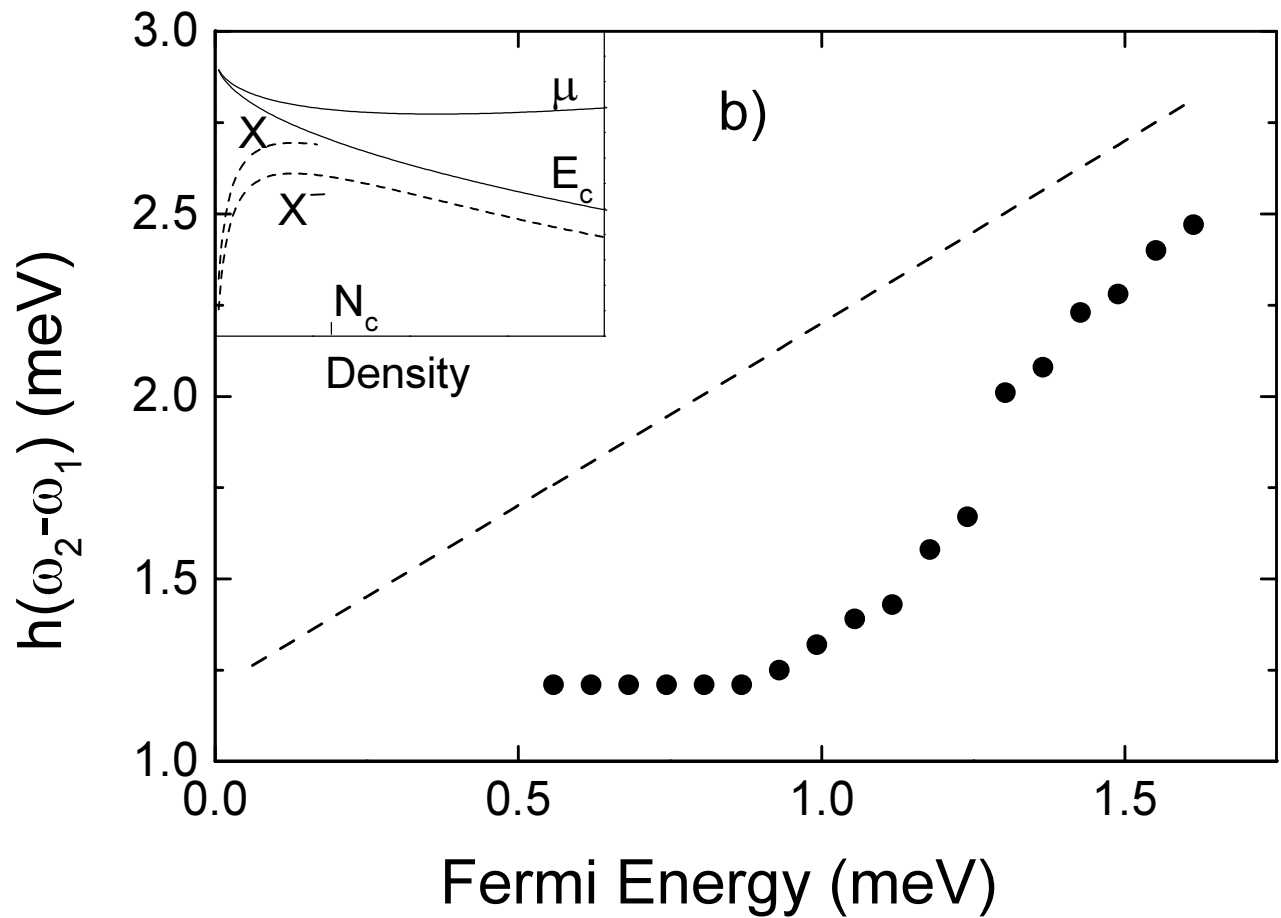
\end{document}